\documentstyle[11pt]{article}

\begin{document} 
\centerline{\bf Chain of nuclear spins system quantum computer taking into account second neighbor
Ising  }
\centerline{\bf spins interaction and  numerical simulation of Shor factorization of N=4}
\vskip2pc
\centerline{G.V. L\'opez and L. Lara}
\centerline{Departamento de F\'{\i}sica, Universidad de Guadalajara}
\centerline{Apartado Postal  4-137, 44410 Guadalajara, Jalisco, M\'exico}
\vskip2pc
\centerline{PACS: 03.67.Lx, 03.65.Ta}
\vskip2cm
\centerline{ABSTRACT}
\vskip1pc\noindent
For a one-dimensional chain of four nuclear spins ($1/2$) and taking into account first and second
neighbor interactions among the spin system,  we make the numerical simulation of Shor
prime factorization algorithm of the integer number $N=4$ to study the influence of the second neighbor 
interaction on the performance of this algorithm. It is shown that the optimum Rabi's frequency to control
the non-resonant effects is dominated by the second neighbor interaction coupling parameter ($J'$), and
that a good Shor quantum factorization is achieved for a ratio of
second to first coupling constant of $J'/J\ge 0.04$.
    
\vfil\eject\noindent 
{\bf 1. Introduction}
\vskip0.5pc\noindent
The polynomial time solution of the prime decomposition of an integer number, given by Shor
factorization algorithm [1] using quantum computation, has triggered the huge amount of work in quantum
computers [2] and quantum information [3] areas in physics. This algorithm has already been demonstrated
for few qubits [4] quantum computers. A qubit is the superposition of two quantum states of the system,
say $|0\rangle$ and $|1\rangle$, $\Psi=C_0|0\rangle+C_1|1\rangle$ such that $|C_0|^2+|C_1|^2=1$, and it
is the basic element to process the information in a quantum computer. The states $|0\rangle$ and
$|1\rangle$ can also be called basic qubits. The
$L$-tensorial product of
$L$-basic-qubits forms a register of length $L$, say $|x\rangle=|i_{L-1},\dots,i_0\rangle$ with $i_j=0,1$
("0" for the ground state and "1" for the exited state).  The set of these states makes up the basis of
the
$2^L$-dimensional Hilbert space where the quantum computer works, and a typical element of this space is
given by
$\Psi=\sum C_x|x\rangle$ with $\sum |C_x|^2=1$. A solid state quantum computer of our particular
interest and which might be developed in a near future is using a one-dimensional chain of nuclear spins
(1/2) which is inside a strong magnetic field in the z-direction and an rf-field in the x-y plane. The
magnetic field in the z-direction determines the state of the nuclear spin, $|0\rangle$ if the nuclear
spin is parallel to this magnetic field and $|1\rangle$ if the nuclear spin is in opposite direction. 
This magnetic field also determines the Zeeman spectrum of the system. The rf-field is used to cause the
desired transitions between the Zeeman levels of the systems. Up to now, this model has been developed
just theoretically and hopefully the technological and experimental part may start in a near future.
However, because the Hamiltonian of this system is well known, very important theoretical studies have
been made [5] which are also important for the general understanding of quantum computation [6]. In this
model, first neighbor interaction among the nuclear spins is considered, and Shor quantum factorization
of the number $N=4$ has been simulated with this model [7].  In this paper, we consider also
second neighbor interaction among the nuclear spins.  Using this interaction, we study Shor factorization
algorithm to  factorize the integer number N=4, developing the proper code to do this. We study the well
performance of this factorization through the fidelity parameter and determine the minimum value of the
second neighbor interaction coupling constant to do this. 
Finally, we also point out the modification caused by the second neighbor interaction to the so called 
$2\pi k$-method, used to suppress non-resonant transitions.

\vskip2pc
\leftline{\bf 2. Equation of Motion}
\vskip1pc\noindent
Consider a one-dimensional chain of four equally spaced nuclear-spins system (spin one half) making an
angle $\cos\theta=1/\sqrt{3}$ with respect the z-component of the magnetic field (chosen in this way to
kill the dipole-dipole interaction between spins) and having an rf-magnetic field in the transversal
plane. The magnetic field is given by
$${\bf B}=(b\cos(\omega t+\varphi), -b \sin(\omega t+\varphi), B(z))\ ,\eqno(1)$$
where $b$, $\omega$ and $\varphi$  are the amplitude, the angular frequency and the phase of the rf-field,
which could be different for different pulses. $B(z)$ is the amplitude of the z-component of the magnetic
field. Thus,
the Hamiltonian of the system up to second neighbor interaction is given by
$$H=-\sum_{k=0}^3{\bf \mu_k}\cdot {\bf B_k}-2J\hbar\sum_{k=0}^2I_k^zI_{k+1}^z
-2J'\hbar\sum_{k=0}^1I_k^zI_{k+2}^z\ ,\eqno(2)$$
where ${\bf\mu_k}$ represents the magnetic moment of the kth-nucleus which is given in terms of the
nuclear spin as ${\bf\mu_k}=\hbar\gamma(I_k^x,I_k^y, I_k^z)$, being $\gamma$ the proton gyromagnetic
ratio. ${\bf B_k}$ represents the magnetic field at the location of the $kth$-spin. The second term at the
right side of (2)  represents the  first neighbor spin interaction, and the third term represents the
second neighbor spin interaction. $J$ and $J'$ are the coupling constants for these interactions. This
Hamiltonian can be written in the  following way
$$H=H_0+W\ ,\eqno(3a)$$
where $H_0$ and $W$ are given by
$$H_0=-\hbar\left\{\sum_{k=0}^3\omega_kI_k^z+2J(I_0^zI_1^z+I_1^zI_2^z+I_2^zI_3^z)
+2J'(I_0^zI_2^z+I_1^zI_3^z)\right\}\eqno(3b)$$
and
$$W=-{\hbar\Omega\over 2}\sum_{k=0}^3\biggl[e^{i\omega t}I_k^++e^{-i\omega t}I_k^-\biggr]\ .\eqno(3c)$$
The term $\omega_k$ represents the Larmore frequency of the kth-spin, $\omega_k=\gamma B(z_k)$. The term
$\Omega$ is the Rabi's frequency, $\Omega=\gamma b$. Finally, the term $I_k^{\pm}=I_k^x\pm iI_k^y$
represents the ascend operator (+) or the descend operator (-). The Hamiltonian $H_0$ is diagonal on the
basis
$\{|i_3i_2i_1i_0\rangle\}$, where $i_j=0,1$ (zero for the ground state and one for the exited state),
$$H_0|i_3i_2i_1i_0\rangle=E_{i_3i_2i_1i_0}|i_3i_2i_1i_0\rangle\ .\eqno(4a)$$
The eigenvalues $E_{i_3i_2i_1i_0}$ are given by
$$E_{i_3i_2i_1i_0}=-{\hbar\over 2}\biggl\{\sum_{k=0}^3(-1)^{i_k}\omega_k+J\sum_{k=0}^2(-1)^{i_k+i_{k+1}}
+J'\sum_{k=0}^1(-1)^{i_k+i_{k+2}}\biggr\}\ .\eqno(4b)$$
The term (3c) of the Hamiltonian allows to have a single spin transitions on the above eigenstates
by choosing the proper resonant frequency. 

\vskip1pc\noindent
To solve the Schr\"odinger equation
$$i\hbar{\partial\Psi\over\partial t}=H\Psi\ ,\eqno(5)$$ 
let us propose a solution of the form
$$\Psi(t)=\sum_{k=0}^{15}C_k(t)|k\rangle\ ,\eqno(6)$$
where we have used decimal notation for the eigenstates in (4a), $H_0|k\rangle=E_k|k\rangle$.
Substituting (6) in (5), multiplying for the bra $\langle m|$, and using the orthogonality relation 
$\langle m|k\rangle=\delta_{mk}$, we get the following equation for the coefficients
$$i\hbar\dot C_m=E_mC_m+\sum_{k=0}^{15}C_k\langle m|W|k\rangle\ \ m=0,\dots,15.\eqno(7)$$
Now, using the following transformation
$$C_m=D_me^{-iE_m t/\hbar}\ ,\eqno(8)$$
the fast oscillation term $E_mC_m$ of Eq. (7) is removed (this is equivalent to go to the interaction
representation), and the following equation is gotten for the coefficients $D_m$
$$i\dot D_m={1\over\hbar}\sum_{k=0}^{15}W_{mk}D_ke^{i\omega_{mk}t}\ ,\eqno(9a)$$
where $W_{mk}$  denotes the matrix elements $\langle m|W|k\rangle$, and $\omega_{mk}$ are defined as
$$\omega_{mk}={E_m-E_k\over\hbar}\ .\eqno(9b)$$
Eq. (9a) represents a set  of 32 real coupling ordinary differential equations which can be solved
numerically, and where $W_{mk}$ are given by
$$W_{mk}=-{\hbar\Omega\over 2}\times(0,\hskip0.5pc {z}\hskip0.5pc\hbox{"or"}\hskip0.5pc {z^*}),\eqno(9c)$$
where $z$ is defined as $z=e^{i(\omega t+\varphi)}$, and $z^*$ is its complex conjugated.
\vskip2pc
\leftline{\bf 3. Shor's factorization algorithm and numerical simulation}
\vskip1pc\noindent
Following Shor's approach to get the factorization of an integer number N, one selects a $L+M$-register
of the form $|x;y\rangle$, where $|x\rangle$ is the input register of length $L$, and $|y\rangle$ is the
valuation register of length $M$. The $|y\rangle$ register will store the values of the periodic function
$y(x)=q^x(mod~N)$, where the integer $q$ is a coprime number of $N$, that is, their grater common divisor
is one ($gcd(q,N)=1$). Thus, starting with the ground state,
$$\Psi_0=|0;0\rangle\ ,\eqno(10)$$
the uniform superposition state is created in the x-register,
$$\Psi_1={1\over 2^{L/2}}\sum_x|x;0\rangle\ ,\eqno(11)$$
where $2^L$ is the number of qubits in the x-register. On the next step, the computation of the function
$y(x)=q^x(mod~N)$ is carried out in the y-register,
$$\Psi_2={1\over 2^{L/2}}\sum_x|x;y(x)\rangle\ .\eqno(12)$$
Then, the discrete Fourier transformation is done in the x-register,
$$\Psi_3={1\over 2^L}\sum_x\sum_ke^{i2\pi kx/2^L}|x;y(x)\rangle\ .\eqno(13)$$
After these steps, one makes the measurement of the state on the x-register, and the probability must be a
peak distribution with peaks separation, $\Delta x$, equal to $\Delta x=2^L/T$, whenever $2^L$ be
divisible by  the period $T$. In this way, one finds the period $T$ of the function $y(x)=q^x(mod~N)$. If
this period is an even number, the factors of $N$ can be computed  finding the greatest common divisor of
$q^{T/2}\pm 1$ and the number $N$ ($gcd(q^{T/2}\pm 1,N)$).
\vskip0.5pc\noindent
Now, for $N=4$, one just needs two-qubits registers in the x-register ($L=2$) and two-qubits registers
in the y-register ($M=2$). The only coprime number is $q=3$, and the function $y(x)=3^x(mod~4)$ has
period $T=2$. Therefore, starting with the function
$$\Psi_0=|00;00\rangle\ ,\eqno(14a)$$
the superposition state is created in the x-register,
$$\Psi_1={1\over 2}\biggl\{|00;00\rangle+|01;00\rangle+|10;00\rangle+|11;00\rangle\biggr\}\ .\eqno(14b)$$
Next, the function $y(x)=3^x(mod~4)$ is valuated on the y-register,
$$\Psi_2={1\over 2}\biggl\{|00;01\rangle+|01;11\rangle+|10;01\rangle+|11;11\rangle\biggr\}\ .\eqno(14c)$$
Then, the discrete Fourier transformation is performed to get (after summation of all terms) the wave 
function
$$\Psi_3={1\over 2}\biggl\{|00;01\rangle+|00;11\rangle+|10;01\rangle+10;11\rangle\biggr\}\ .\eqno(14d)$$
The measurement on the x-register give us the states $|00\rangle$ or $10\rangle$ ($x=0$ or $x=2$). So, one
has
$\Delta x=2$, and the period of our function is $T=2^2/2=2$. Finally, the factors of $N=4$ ($4=2\cdot
2$) are obtained from $gcd(3^{T/2}-1,4)=2$.
\vskip0.5pc\noindent
To make the numerical simulation of this algorithm, we have chosen the following  parameters
in units of $2\pi\times$Mhz,
$$\omega_0=100\ ,\ \omega_1=200\ ,\ \omega_2=400\ ,\ \omega_3=800,\ J=10\ ,\ J'=0.4\ ,\
\Omega=0.1\ .\eqno(15)$$ 
These parameters were chosen in this way to have a clear separation on the Zeeman spectrum and to have a
good definition for the resonant transitions in our numerical simulation. This does not imply a
restriction on our simulations since our main results are applicable also to the current design [6]. Now,
starting with the ground state of the system,
$|0000\rangle$, we create a superposition state in the x-register using three $\pi/2$-pulses with zero
phases and with resonant frequencies
$\omega_{0,4}$,
$\omega_{0,8}$ and $\omega_{4,12}$. The valuation of the function $y(x)=3^x(mod~4)$ in the y-register is
carried out with four $\pi$-pulses with zero phases and with the resonant frequencies $\omega_{0,1}$,
$\omega_{4,5}$, $\omega_{5,7}$ and $\omega_{13,15}$. Finally, the discrete Fourier transformation in the
x-register is gotten through five $\pi$-pulses with zero phases and with the frequencies $\omega_{6,7}$,
$\omega_{2,6}$, $\omega_{2,3}$, $\omega_{14,15}$, and $\omega_{11,15}$. The transitions involved in the
algorithm are shown in Fig. 1. Note that at the end of Shor's algorithm, one must get the following
probabilities
$$|C_1|^2=|C_3|^2=|C_9|^2=|C_{11}|^2=1/4\ ,\eqno(16)$$
according with our final wave function (14d), note from (8) that $|C_k|=|D_k|$. Fig. 2 shows the
behavior of the probabilities $|C_k|^2$
 during the entire Shor's algorithm. Fig. 2a shows the formation of the
superposition state (14b), starting from the ground state (14a). Fig. 2b shows the valuation of the
function $y(x)=3^x(mod~4)$ at the end of four $\pi$-pulses, wave function (14c). Fig. 2c shows the
formation of the wave function (14d) at the end of five $\pi$-pulses. To better illustrate what is
happening during the Shor's algorithm,  we calculated the expected values of the z-component of the spin
of the system for each qubit. These expected values are given by
$$\langle I_0^z\rangle={1\over 2}\sum_{k=0}^{15} (-1)^k|C_k(t)|^2\ ,\eqno(17a)$$
\begin{eqnarray*}
\langle I_1^z\rangle&=& {1\over 2}\Biggl\{|C_0|^2+|C_1|^2-|C_2|^2-|C_3|^2+|C_4|^2+|C_5|^2-|C_6|^2-|C_7|^2
\\ \\
& &+|C_8|^2+|C_9|^2-|C_{10}|^2-|C_{11}|^2+|C_{12}|^2+|C_{13}|^2-|C_{14}|^2-|C_{15}|^2\Biggr\}\ ,
\end{eqnarray*}
$$\eqno(17b)$$
\begin{eqnarray*}
\langle I_2^z\rangle&=& {1\over 2}\Biggl\{|C_0|^2+|C_1|^2+|C_2|^2+|C_3|^2-|C_4|^2-|C_5|^2-|C_6|^2-|C_7|^2
\\ \\
& &+|C_8|^2+|C_9|^2+|C_{10}|^2+|C_{11}|^2-|C_{12}|^2-|C_{13}|^2-|C_{14}|^2-|C_{15}|^2\Biggr\}\ ,
\end{eqnarray*}
$$\eqno(17c)$$
and
$$\langle I_3^z\rangle={1\over 2}\sum_{k=0}^7|C_k|^2-{1\over 2}\sum_{k=8}^{15}|C_k|^2\ .\eqno(17d)$$
Fig. 3a shows these expected values during the formation of the superposition state on the x-register.
Fig. 3b shows their behavior during the valuation of the function $y(x)=3^x(mod~4)$ on the y-register,
and Fig. 3c shows their behavior during the creation of the discrete Fourier transformation. Of course,
this observed behavior is the behavior that one could have expected from the Shor's algorithm.  Fig. 4a
shows the probabilities of the expected four-qubits registers at the end of Shor's algorithm (wave
function (14d)), and Fig. 4b shows the probabilities of the non-resonant states. 
\vskip0.5pc\noindent
To determine the value of the second neighbor coupling constant needed to have a good reproduction of the
Shor's algorithm, we calculate the fidelity parameter [8],
$$F=\langle\Psi_{expected}|\Psi\rangle\ ,\eqno(18)$$
where $\Psi_{expected}$ is the wave function (14d), and $\Psi$ is the resulting wave function from our
computer simulation. This is done as a function of the ratio $J'/J$ (second  to first neighbor coupling
constant interactions). Fig. 5 shows our results of the calculation of this parameter as a function of
$J'/J$. As one can see, for a value of $J'/J\ge 0.04$, one can say that we have already a very good
reproduction of the Shor factorization algorithm. Of course, this does not mean that second neighbor
interaction is needed to implement Shor quantum algorithm since, if $J'=0$, one can just change the
protocol (resonant pulses) to get this algorithm. Rather, what this result means is that, in the case of
having second neighbor interaction ($J'\not=0$), the optimum performance of our Shor quantum algorithm is
gotten for the above ratio.

\vskip2pc\noindent
\leftline{\bf 4. Second neighbor interaction and the $2\pi k$-method}
\vskip0.5pc\noindent
One of the important results from the consideration of first neighbor interaction and the selection of
the parameter as $J/\Delta\omega\ll 1$ and $\Omega/\Delta\omega\ll 1$ is the possibility of choosing the
Rabi's frequency
$\Omega$ in such a way that the non-resonant effects are eliminated. This procedure is called the $2\pi
k$-method [9], and this Rabi's frequency for a $\pi$-pulse is chosen as $\Omega=|\Delta|/\sqrt{4k^2-1}$,
where $k$ is an integer number, and $\Delta$ is the detuning parameter 
between the states $|p\rangle$ and $|m\rangle$, $\Delta=(E_p-E_m)/\hbar-\omega$ being $\omega$ the
electromagnetic  resonant frequency. This detuning parameter is proportional to the first neighbor
coupling constant $J$. Let us see how this detuning parameter is modified due to second neighbor
interaction. Assuming that the states
$|p\rangle$ and
$|m\rangle$ are are the only ones involved in the dynamics, from Eq. (9a), one has
$$i\dot D_m={W_{mp}\over\hbar}D_pe^{i\omega_{mp}t}\ ,\hskip1.5pc\hbox{and}\hskip2.5pc i\dot
D_p={W_{pm}\over\hbar}D_m e^{i\omega_{pm}t}\ .\eqno(19)$$
Thus, given the initial conditions $D_p(0)=C_p(0)$ and $D_m(0)=C_m(0)$, the solution is readily given by
$$D_p(t)=\Biggl\{C_p(0)\biggl[\cos{\Omega_et\over 2}-i{\Delta\over\Omega_e}\sin{\Omega_et\over 2}\biggr]+
i{\Omega C_m(0)\over\Omega_e}\sin{\Omega_e t\over 2}\Biggr\}e^{i\Delta t\over 2}\eqno(20a)$$
and
$$D_m(t)=\Biggl\{C_m(0)\biggl[\cos{\Omega_et\over 2}-i{\Delta\over\Omega_e}\sin{\Omega_et\over 2}\biggr]+
i{\Omega C_p(0)\over\Omega_e}\sin{\Omega_e t\over 2}\Biggr\}e^{-i\Delta t\over 2}\ ,\eqno(20b)$$
where $\Omega_e$ is defined as $\Omega_e=\sqrt{\Omega^2+\Delta^2}$. So, for example, at the end of a
$\pi$-pulse ($t=\tau=\pi/\Omega$), the argument of the periodic functions in (20a) and (20b) is given by 
$\Omega_e\pi/2\Omega$. If one makes this term to be equal to any multiple of $\pi$,  one can get rid of
the non-resonant terms since the solutions are given by
$$D_p(\tau)=(-1)^kC_p(0)e^{i\Delta\pi/2\Omega}\ ,\hskip0.5pc\hbox{and}\hskip0.5pc 
D_m(\tau)=(-1)^kC_m(0)e^{-i\Delta\pi/2\Omega}\ .$$
The Rabi's frequency obtained from the condition $\Omega_e\pi/2\Omega=k\pi$ is given by
$\Omega=|\Delta|/\sqrt{4k^2-1}$, and this is the so called $2\pi k$-method.
\vskip0.5pc\noindent
Now, let us select a resonant transition containing the
Larmore frequency $\omega_0$. These frequencies could be $\omega_0+J+J'$, $\omega_0-J+J'$,
$\omega_0-J-J'$ or $\omega_0+J-J'$ which correspond to the transitions (decimal notation)
$|0\rangle\leftrightarrow |1\rangle~(|10\rangle
\leftrightarrow |11\rangle)$, $|2\rangle\leftrightarrow|3\rangle$, $|6\rangle\leftrightarrow|7\rangle$, 
and $|2\rangle\leftrightarrow|3\rangle$. So, all of these states are pertubated during a pulse, and the
frequency difference $\Delta$ may have the values $2J$, $2J'$, $2J+2J'$, or $2J-2J'$. Choosing other
Larmore frequencies, the additional values for the detuning parameter are $4J$, and $4J+2J'$. 
Thus, denoting by
$\Omega_{\Delta}^{(k)}$ the Rabi's frequency selected by this method,
$$\Omega_{\Delta}^{(k)}={|\Delta|\over\sqrt{4 k^2-1}}\ ,\eqno(21)$$
there are five possible values for $\Delta$ which are $4J+2J'$, $4J$, $2J+2J'$, $2J$ and $2J'$.
\vskip0,5pc\noindent
To see the dependence of the Shor's algorithm with respect the Rabi's frequency, we use again the
fidelity parameter parameter (18). Using the same values for our parameters as (15) but $\Omega$, Fig. 6
shows the fidelity parameter as a function of the Rabi's frequency. The lines $L1$, $L2$ and $L3$ mark
the $\Omega's$ values  where this fidelity has peaks. These peaks correspond to the following $2\pi
k$-method Rabi frequencies

$$\Omega_{4J+2J'}^{(252)}\approx
\Omega_{4J}^{(250)}\approx
\Omega_{2J+2J'}^{(129)}\approx
\Omega_{2J}^{(124)}\approx
\Omega_{2J'}^{(5)}=0.080403\ ,$$
$$\Omega_{4J+2J'}^{(202)}\approx
\Omega_{4J}^{(198)}\approx
\Omega_{2J+2J'}^{(103)}\approx
\Omega_{2J}^{(98)}\approx
\Omega_{2J'}^{(4)}=0.100791\ ,$$
and
$$\Omega_{4J+2J'}^{(150)}\approx
\Omega_{4J}^{(147)}\approx
\Omega_{2J+2J'}^{(76)}\approx
\Omega_{2J}^{(73)}\approx
\Omega_{2J'}^{(3)}=0.135225\ .$$
As one can see in Fig. 7, where we have plotted $\Omega_{\Delta}^{(k)}$ (for the detuning values
mentioned above) and where the lines $L1$, $L2$ and $L3$ have been drawn, around these lines there are
several other values of $\Omega_{4J+2J'}^{(k)}$, $\Omega_{4J}^{(k)}$, $\Omega_{2J+2J'}^{(k)}$ and 
$\Omega_{2J}^{(k)}$ which, in principle, should cause a peak in the fidelity parameter (because they
belong to the $2\pi k$-method). However, they do not appear at all on Fig. 6. This means that the peaks
values on the fidelity parameter are fully dominated by the second neighbor coupling interaction parameter
($J'$).  
\vskip3pc\noindent
\leftline{\bf 5. Conclusions}
\vskip0.5pc\noindent
For a  one-dimensional chain of nuclear spins (one half) system quantum computer, we have considered
first and second neighbor interactions, and we have study the effects of second neighbor interaction on
Shor quantum factorization algorithm of the number $N=4$. We have shown that a good factorization
algorithm can be gotten if the ratio of second to first neighbor interaction constants is chosen such
that $J'/J\ge 0.04$. We have shown also that the application of $2\pi k$-method to eliminate non-resonant
transitions  is not so simple since the detuning factor
varies with both parameters $J$ and $J'$ (first and second neighbor coupling interactions). However, the
peaks on the fidelity parameter are dominated by the second neighbor coupling parameter. In other words,
if there exists second neighbor interaction in a chain of nuclear spin system quantum computer, this
interaction may dominate the $2\pi k$-method to control non-resonant effects.

\vskip5pc\noindent
\leftline{\bf Acknowledgements}
\vskip0.5pc\noindent
 This work was supported by SEP under the contract PROMEP/103.5/04/1911 and the University of Guadalajara.
\vfil\eject
\leftline{\bf Figure Captions}
\vskip1pc\noindent
Fig. 1 Energy levels and resonant frequencies used within the algorithm.
\vskip0.5pc\noindent
Fig. 2 Probabilities $|C_k|^2: (k)$. (a) Formation of superposition state in the x-register, wave
function (14b).  (b) Formation of the wave function (14c). (c) Formation the wave function (14d).
\vskip0.5pc\noindent
Fig. 3 Expected values $\langle I_k^z\rangle$: (k=0,1,2,3). (a) During formation of wave function (14b).
(b) During formation of the wave function (14c). (c) During formation of the wave function (14d). 
\vskip0.5pc\noindent
Fig. 4 Probabilities $|C_k|^2$. (a) For the expected states $k=1,3,5,7$. (b) For the non-resonant
states $k=0,2,4,6,8,9,10,11,12,13,14,15$.
\vskip0.5pc\noindent
Fig. 5 Fidelity parameter as a function of $J'/J$.
\vskip0.5pc\noindent
Fig. 6 Fidelity parameter as a function of $\Omega$.
\vskip0.5pc\noindent
Fig. 7 Rabi frequency $\Omega_{\Delta}^{(k)}$ as a function of $k$ for 
$\Delta=4J+2J'$ (1), $\Delta=4J$ (2), $\Delta=2J+2J'$ (3), $\Delta=2J$ (4), $\Delta=2J'$ (5). Lines
$L1$, $L2$ and $L3$ correspond to Fig. 6.
\vfil\eject
\leftline{\bf References}
\obeylines{
1. P.W. Shor, {\it Proc. of the 35th Annual Symposium on the Foundation
\quad of the Computer Science}, IEEE, Computer Society Press, N.Y. 1994, 124.
\quad P.W. Shor, Phys. Rev. A {\bf 52} (1995) R2493.
2. C. Williams and S. Clearwater, {\it Exploration in Quantum Computing},
\quad Springer-Verlag, Berlin, 1995.
3. M.A. Nielsen and I.L. Chuang, {\it Quantum Computation and Quantum 
\quad Information}, Cambridge University Press, 2000.
4. D. Boshi, S. Branca, F.D. Martini, L. Hardy, and S. Popescu
\quad Phys. Rev. Lett., {\bf 80} (1998) 1121.
\quad C.H. Bennett and G. Brassard, {\it Proc. IEEE international Conference on 
\quad Computers, Systems, and Signal Processing}, N.Y. (1984) 175.
\quad I.L. Chuang, N.Gershenfeld, M.G. Kubinec, and D.W. Lung
\quad Proc. R. Soc. London A, {\bf 454} (1998) 447.
\quad I.L. Chuang, N. Gershenfeld, and M.G. Kubinec
\quad Phys. Rev. Lett., {\bf 18} (1998) 3408.
\quad I.L. Chuang, L.M.K. Vandersypen, X.L. Zhou, D.W. Leung, and S. Lloyd,
\quad Nature, {\bf 393} (1998) 143.
\quad P.Domokos, J.M. Raimond, M. Brune, and S. Haroche,
\quad Phys. Rev. Lett., {\bf 52} (1995) 3554.
\quad J.Q. You, Y. Nakamura, F.Nori, Phys. Rev. Lett.,{\bf 91} (2002) 197902.
\quad L.M.K. Vandersypen, M. Steffen, G. Breyta, C.S. Yannoni, M.H. Sherwood
\quad and I.L. Chuang, Nature, {\bf 414} (2001) 883.
5. G.P. Berman, G.D. Doolen, D.D. Holm and V.I. Tsifrinovich,
\quad Phys. Lett. A {\bf 193} (1994) 444.
6. G.P. Berman, G.D. Doolen, D.I. Kamenev, G.V. L\'opez, and
\quad  V.I. Tsifrinovich, Phys. Rev. A {\bf 6106} (2000) 2305.
7. G.P. Berman, G.D. Doolen, G.V. L\'opez, and V.I. Tsifrinovich
\quad quant-ph/9909027  (1999).

8. A. Peres, Phys. Rev. A {\bf 30} (1984) 1610.
\quad B. Schumacher, Phys. Rev. A.,{\bf 51} (1995) 2738.
9. G.P. Berman, G.D. Doolen, D.I. Kamenev, G.V. L\'opez and V.I. Tsifrinovich
\quad Contemporary Mathematics, {\bf 305} (2002)  13.
}

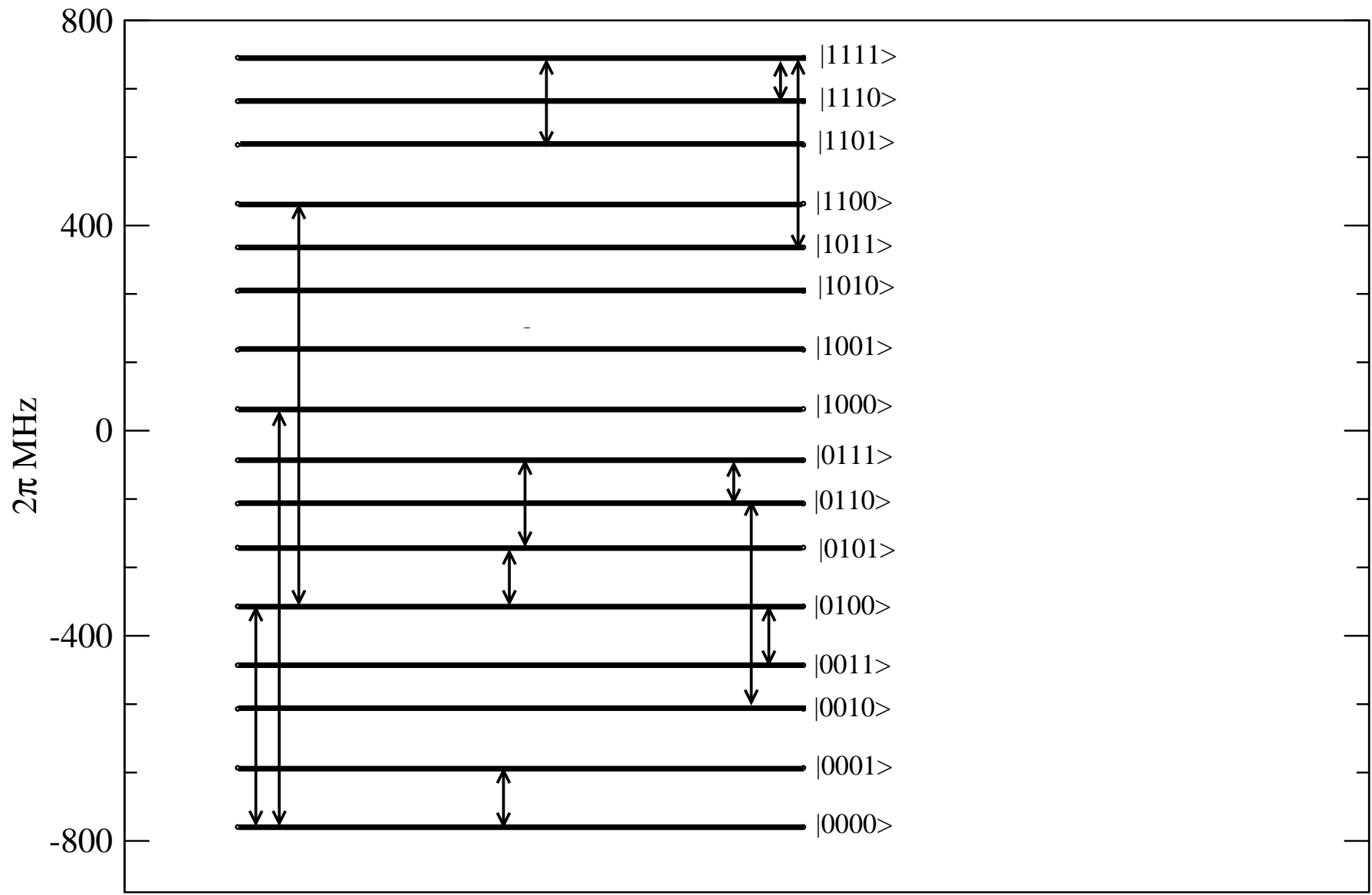
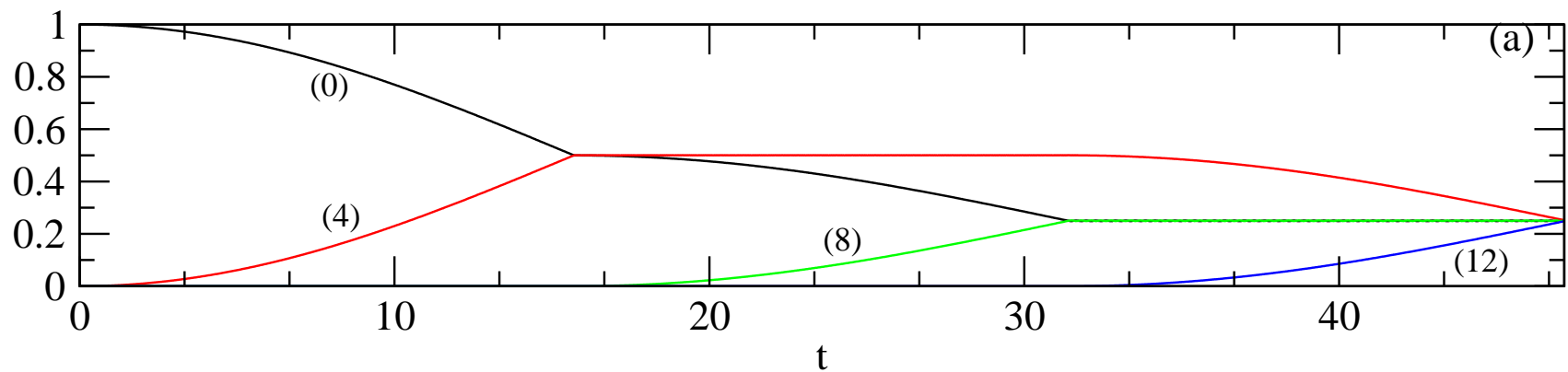
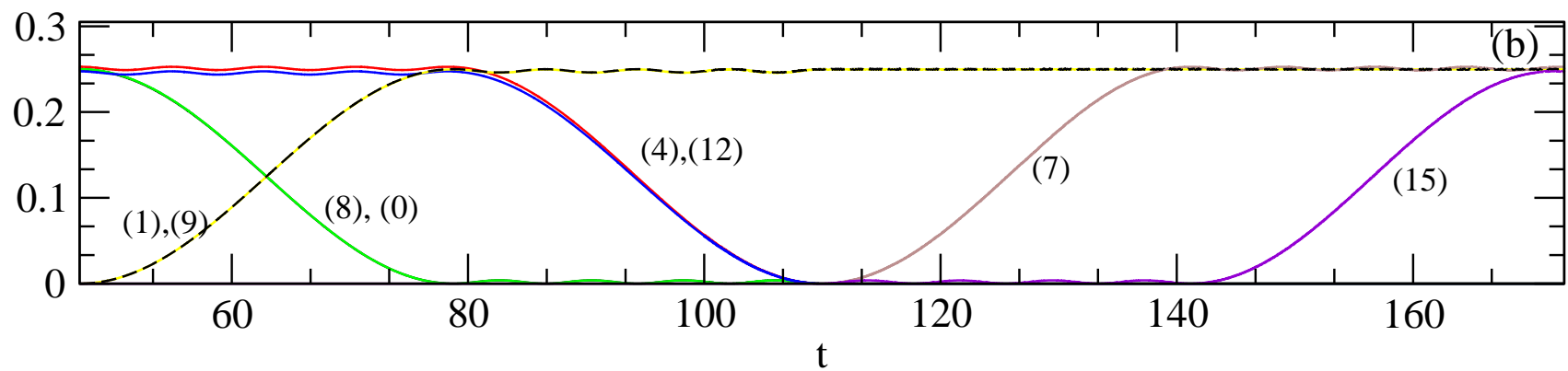
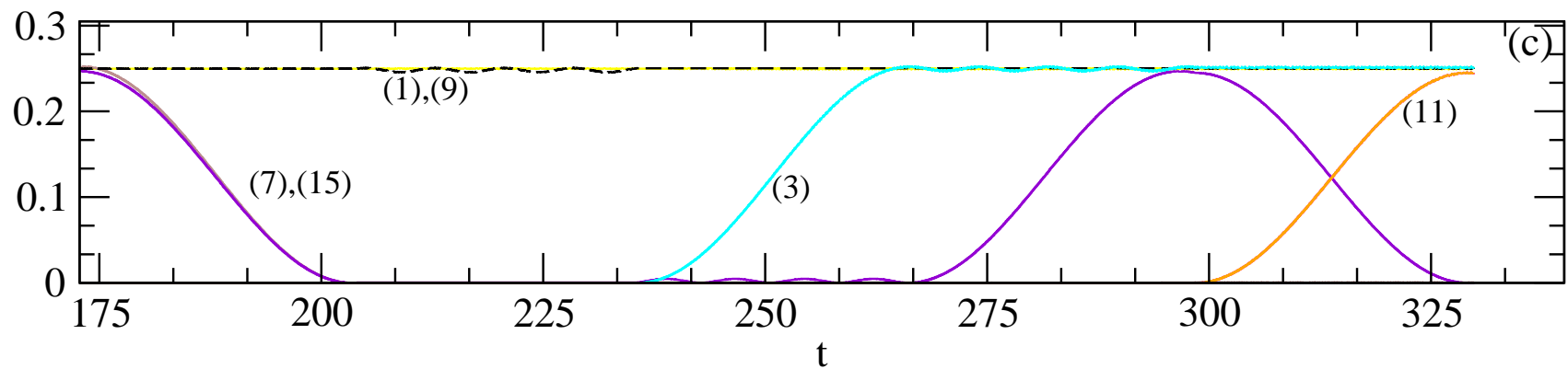
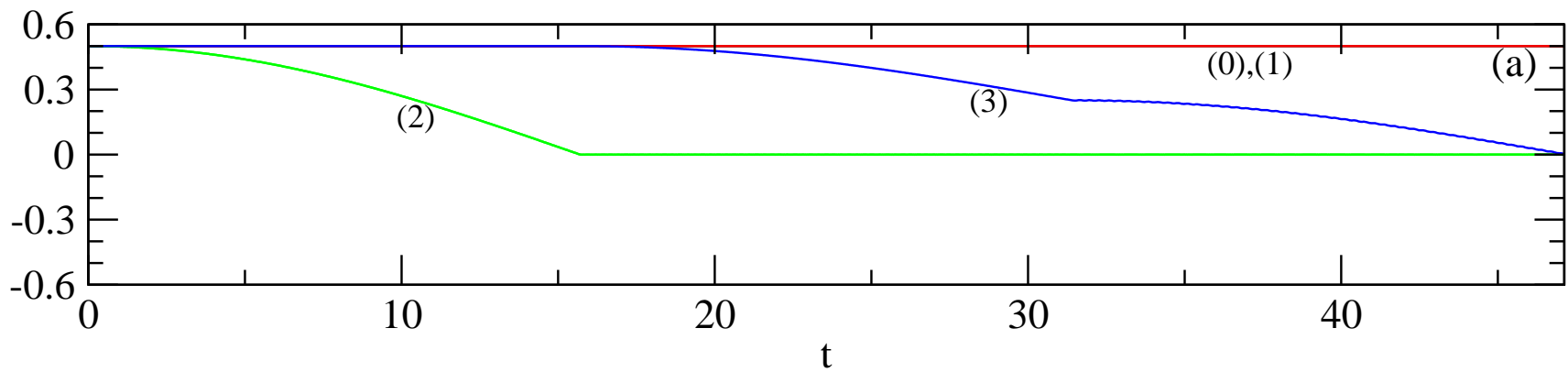
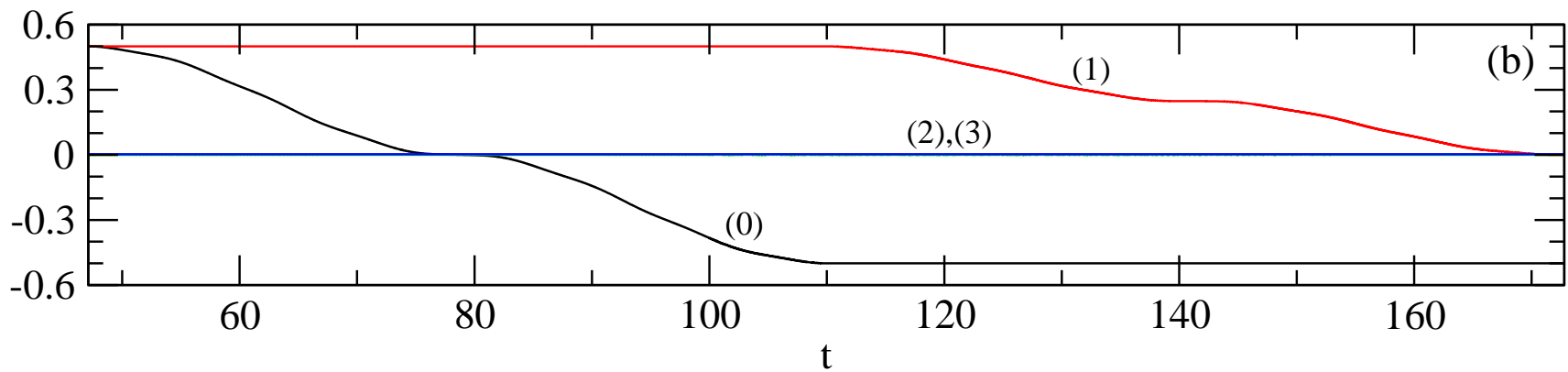
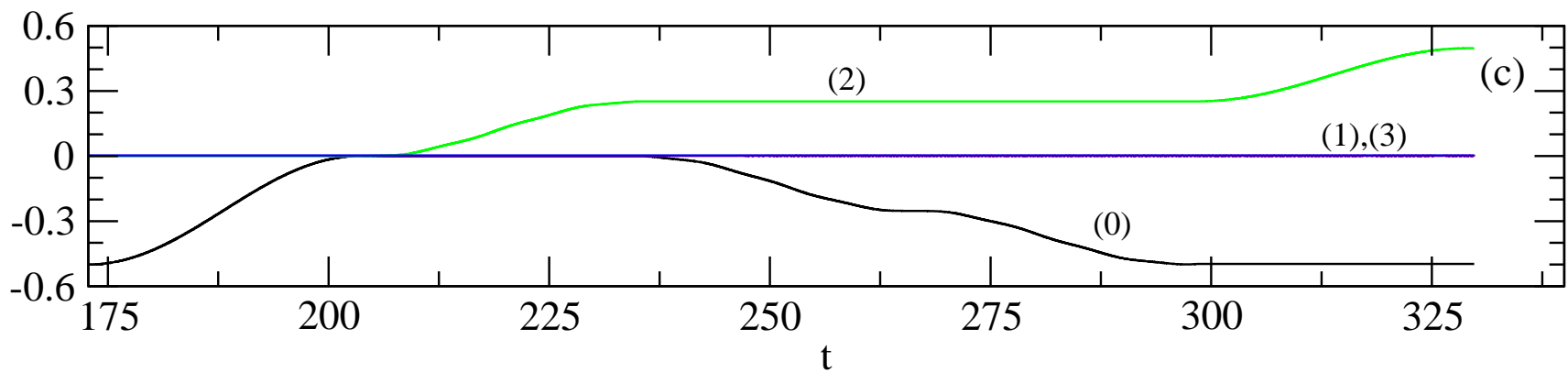
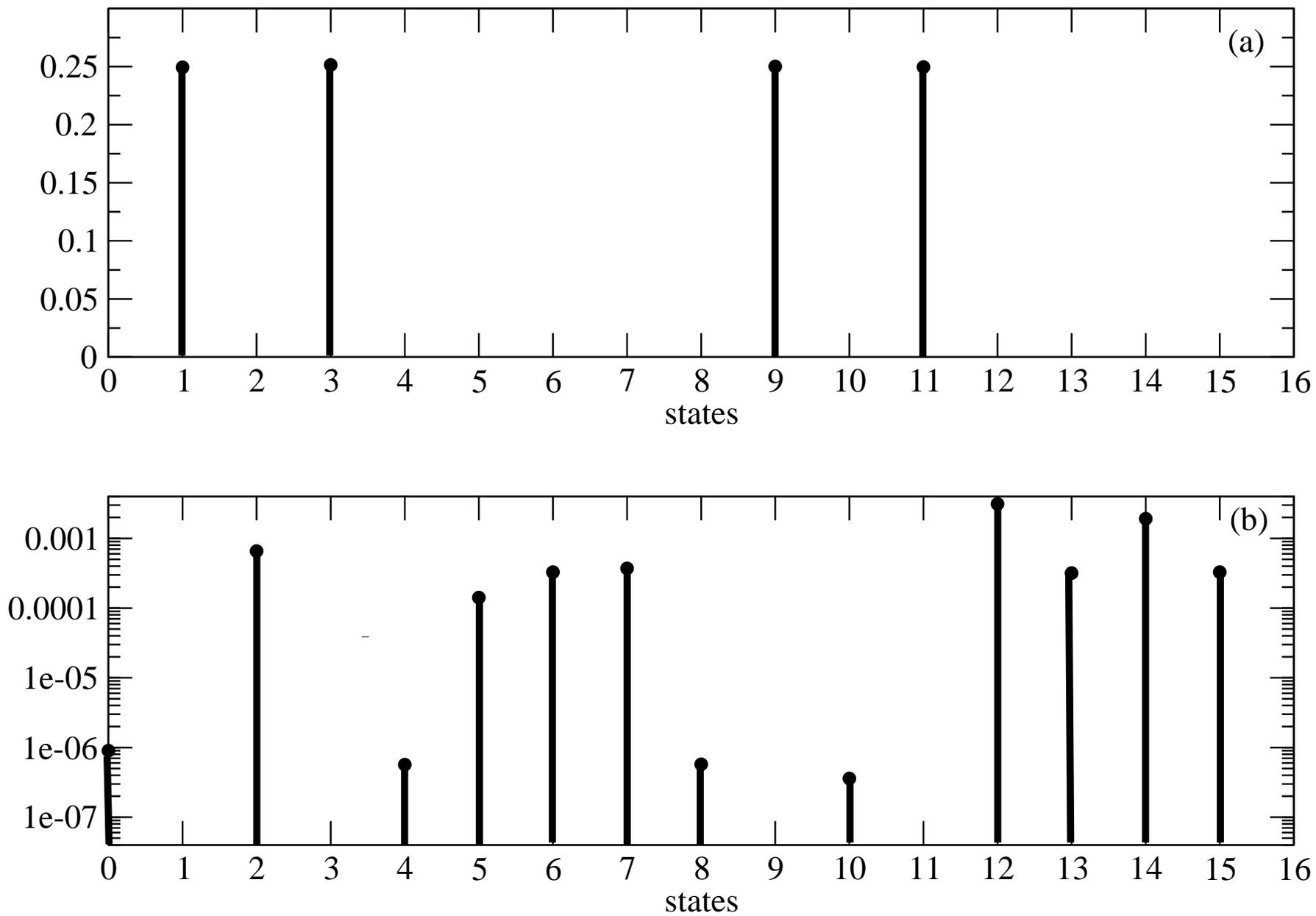
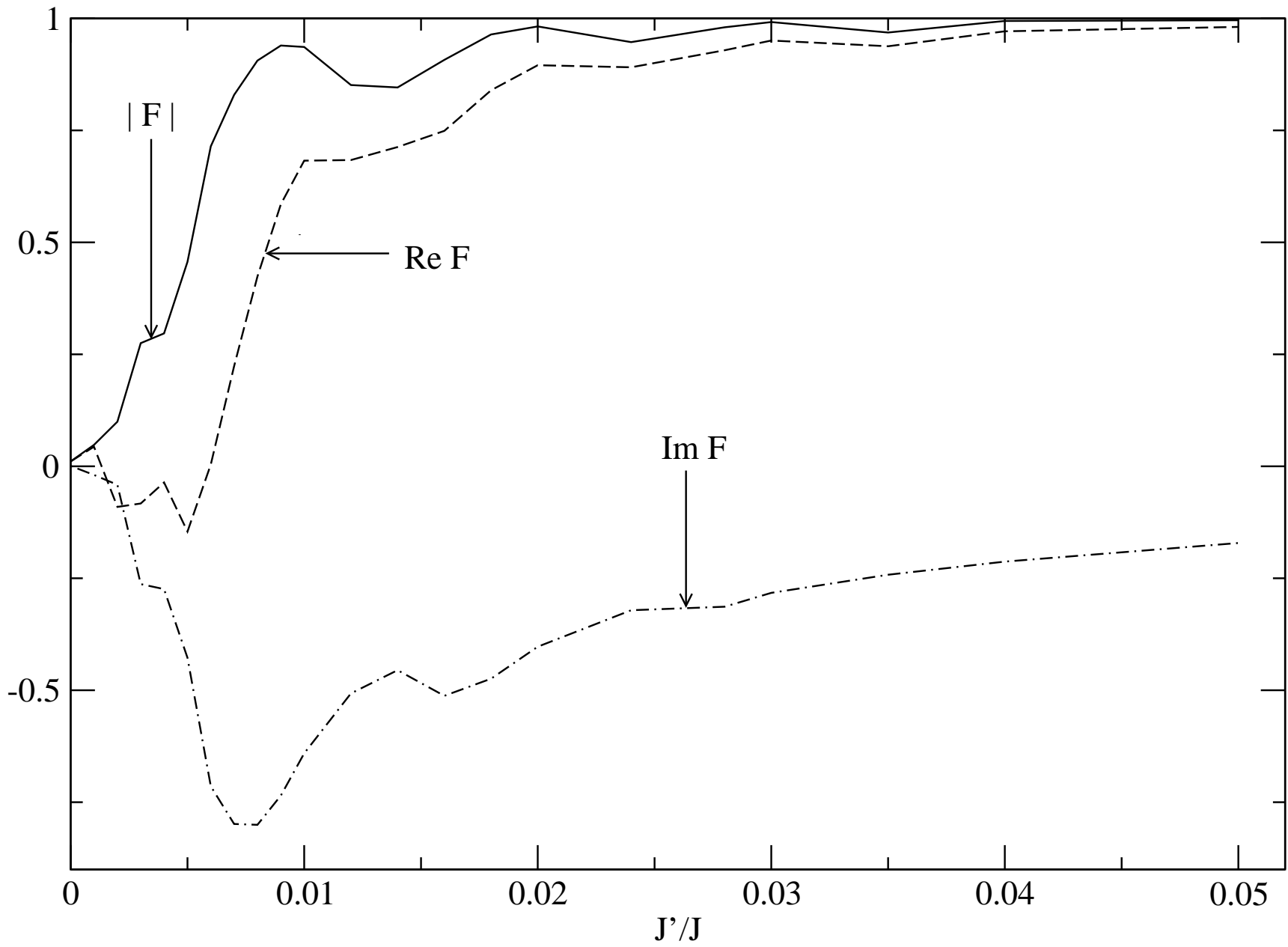
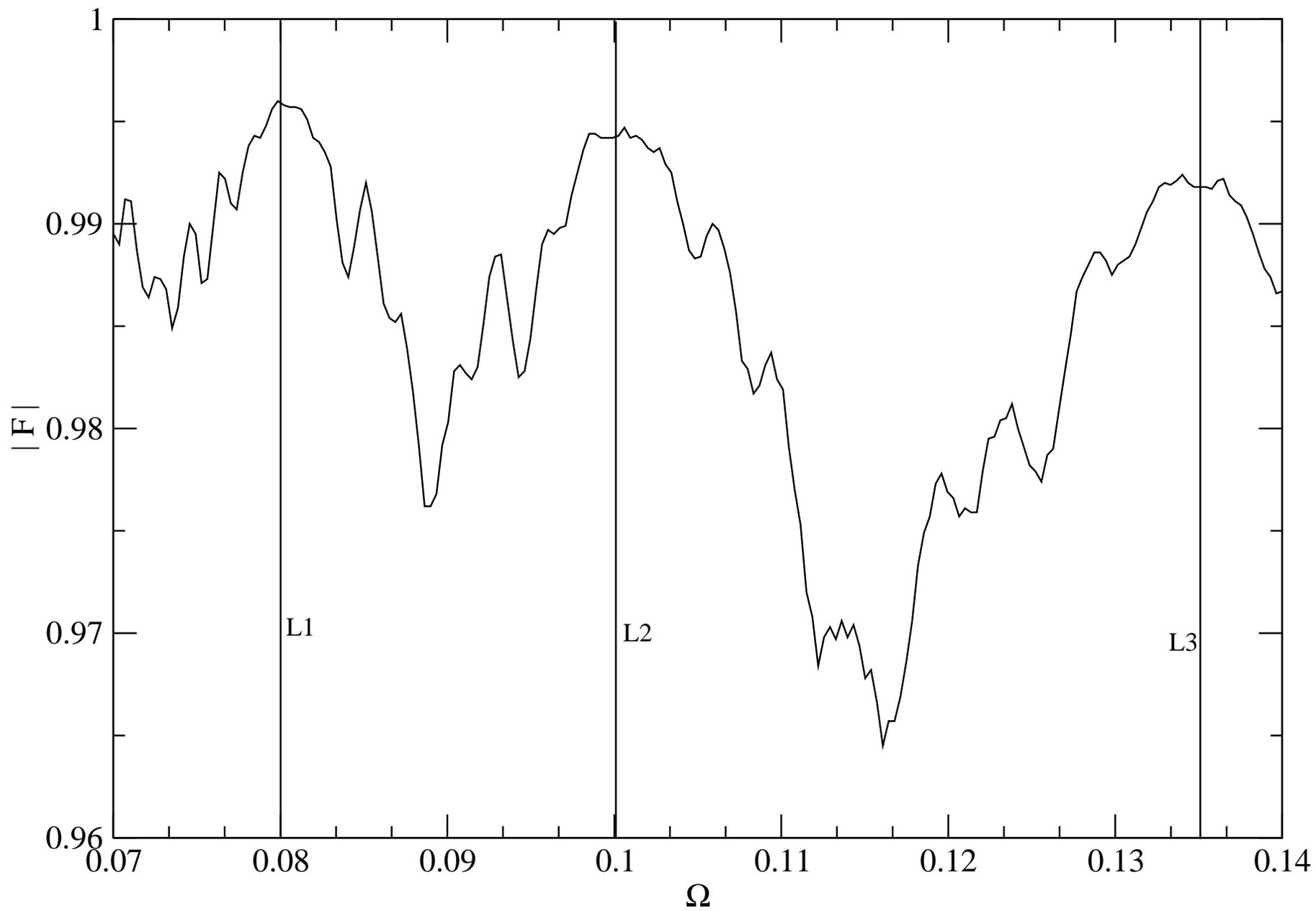
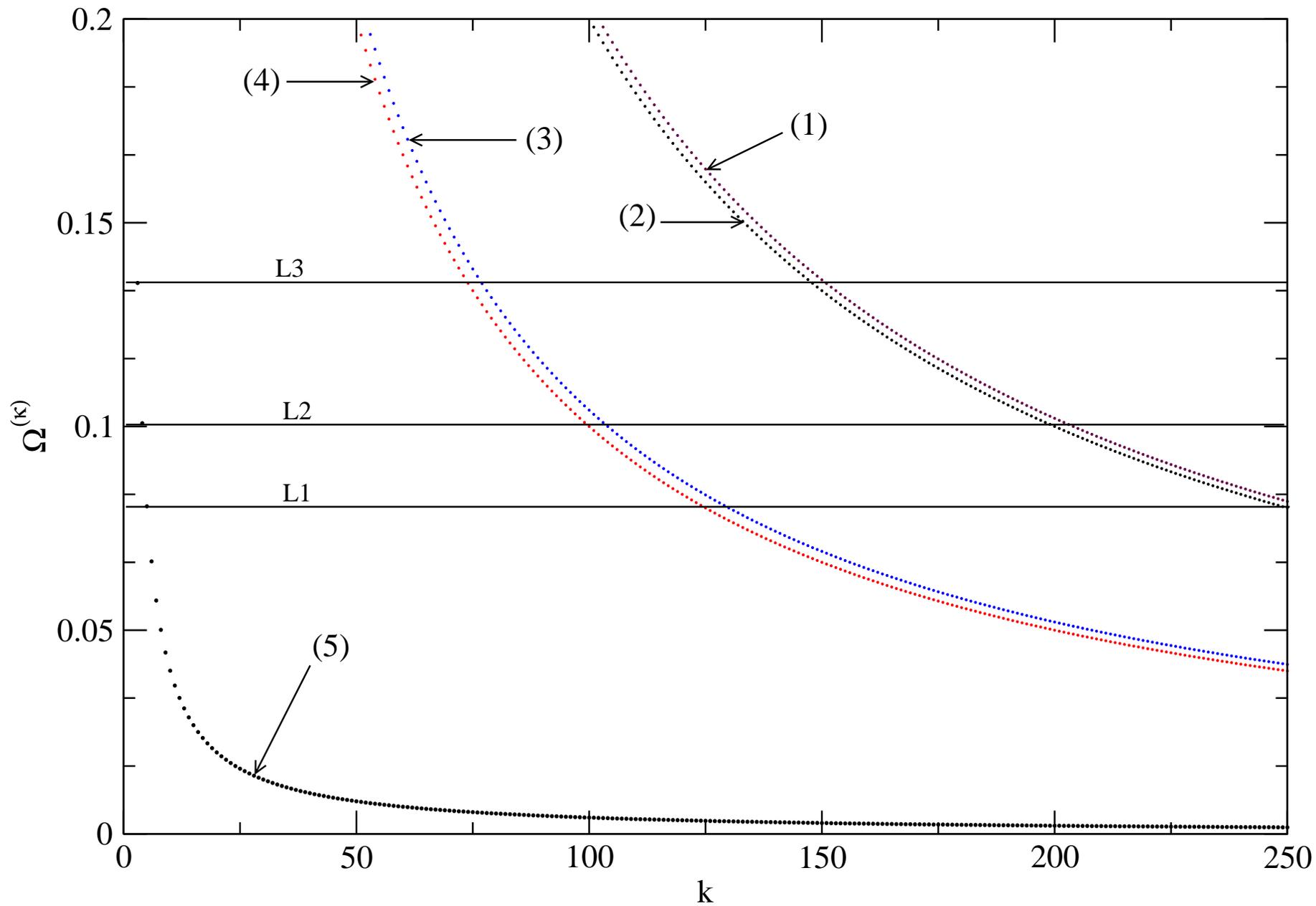
\end{document}